\documentclass[a4paper,twocolumn]{esa-template/esapub2005}

\usepackage[utf8]{inputenc}
\usepackage{times}

\usepackage{hyperxmp}

\usepackage[
    type={CC},
    modifier={by},
    version={4.0},
    ]{doclicense}

\usepackage[
    backend=biber,
    natbib=true,
    url=true, 
    doi=true,
    eprint=false,
    sorting=none
]{biblatex}

\usepackage[breaklinks]{hyperref}
\usepackage{breakurl}

\hypersetup{
    colorlinks,
    linkcolor={red!50!black},
    citecolor={blue!50!black},
    urlcolor={blue!80!black},
    pdftitle={Electronics and Sensor Subsystem Design for Daedalus 2 on REXUS 29: An Autorotation Probe for Sub-Orbital Re-Entry},
    pdfauthor={Jan M. Wolf and Lennart Werner and Philip Bergmann and Clemens Riegler and Frederik Dunschen},
    pdfkeywords={Robust Electronics, Re-Entry, Autorotation, REXUS 29}
}

\usepackage{graphicx}
\usepackage{siunitx}
\DeclareSIUnit\bar{bar}

\usepackage{chemformula}
\usepackage{tabularx}
\usepackage{booktabs}

\usepackage[USenglish]{isodate}

\usepackage{layouts}

\usepackage[disable]{todonotes}
\usepackage[normalem]{ulem}

\usepackage{tikz}
\usepackage{tikz-3dplot}
\usepackage[european,nooldvoltagedirection,fetbodydiode]{circuitikz}
\ctikzset{bipoles/length=1.1cm}
\ctikzset{resistor = european}

\newcommand{\D}{} 
\protected\def\D #1{Daedalus~#1}

\title{ELECTRONICS AND SENSOR SUBSYSTEM DESIGN FOR DAEDALUS 2 ON REXUS 29: AN AUTOROTATION PROBE FOR SUB-ORBITAL RE-ENTRY}
\author{Jan M. Wolf}
\author{Lennart Werner}
\author{Philip Bergmann}
\author{Clemens Riegler}
\author{Frederik Dunschen}
\affil{WüSpace e. V. , Emil-Fischer-Str. 32, 97074 Würzburg, Germany\\
jan.wolf@wuespace.de; lennart.werner@wuespace.de; philip.bergmann@wuespace.de;\\
clemens.riegler@wuespace.de; frederik.dunschen@wuespace.de}

\makeatletter
\apptocmd{\@maketitle}{
	\vspace{-1cm}
	\centering
	Except where otherwise noted, this work is licensed under \doclicenseNameRef \doclicenseIcon
	\vspace{1cm}}
\makeatother

\begin{document}

	\maketitle

	\begin{abstract}
		The \D2 mission aboard REXUS 29 is a technology demonstrator for an alternative descent
mechanism for very high altitude drops based on auto-rotation.
It consists of two probes that are ejected from a sounding rocket at an altitude
of about \SI{80}{\kilo\meter} and decelerate to a soft landing using only a
passive rotor with pitch control.

This type of autonomous, scientific experiment poses great challenges upon the electronics
subsystem, which include mechanical stress, power system reliability, sensor redundancy, subsystem
communication, and development procedures.

Based on the data gathered in \D1 \cite{D1ESAPAC} multiple new approaches were
developed to fulfill these requirements, such as redundant communication links,
mechanical decoupling of PCBs and fault-tolerant power source selection.
	\end{abstract}

	\section{Introduction}
	The experiment will fly as a payload on an Improved Orion sounding rocket as
part of the REXUS/BEXUS program and will test a novel form of re-entry and
descent control without parachutes,
adding to the knowledge base of rotors in space applications, especially
regarding the descent of interplanetary probes.

The use of autorotation has already been conceptually demonstrated by the
predecessor project \D1,
which operated completely passively.
In contrast, \D2 decouples the rotor from the main body and uses an active
control system for sink rate and rotor rotation speed by changing the pitch of
the rotor blades,
which poses great challenges for the electronics and software implementation.

The implementation details of the mechanical setup, with focus on the complex
blade tilting mechanism, were already covered in \cite{D2SSEA}.

\begin{figure}
	\centering
	\includegraphics[width=\columnwidth]{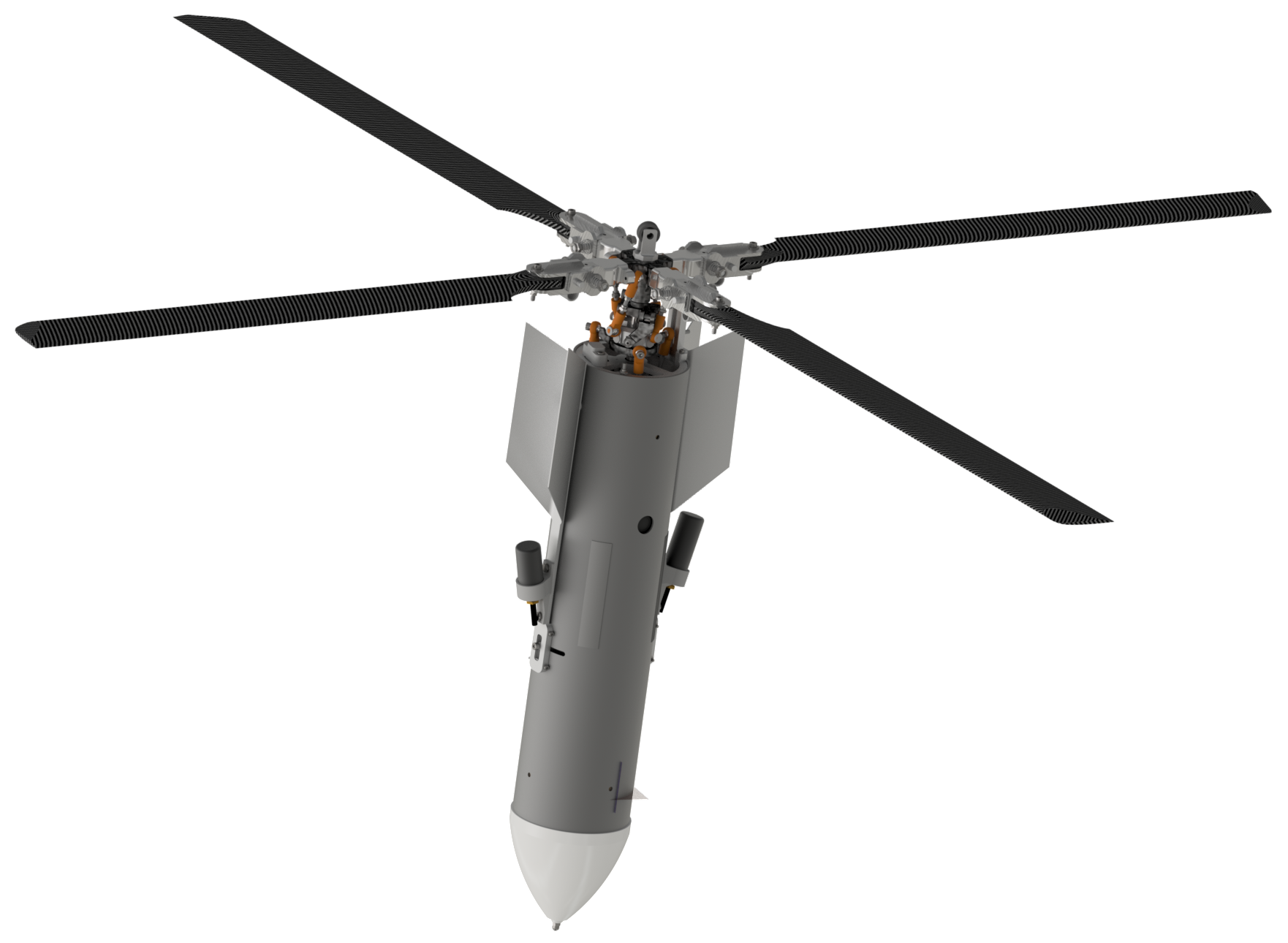}
	\caption{Render of Spaceseed}
	\label{fig:spacesed}
\end{figure}

	\section{System Topology}
	\label{sec:system-topology}
\begin{figure*}
	\centering
	\includegraphics[width=0.95\textwidth]{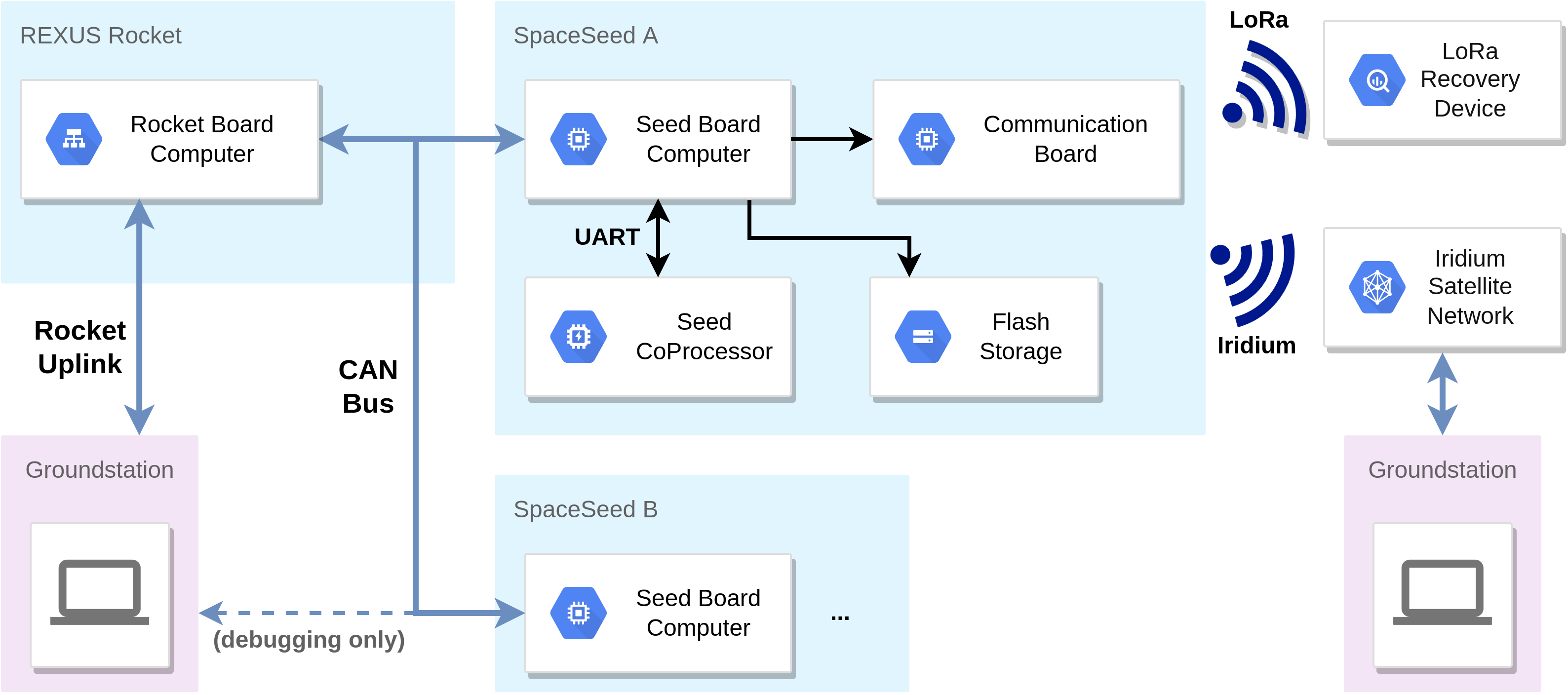}
	\caption{System Topology}
	\label{fig:sys-topo}
\end{figure*}

	The \D2 experiment consists, as most space missions do, of a space,
a ground and a user segment.
The user segment covers post-flight data analysis and will not be covered here
in detail.
It receives the recorded and recovered data after the flight from the ground
segment.
A schematic overview of the space and ground segment is presented in
Fig.~\ref{fig:sys-topo}.

The rocket board computer (RBC) and the Spaceseeds with their Seed Board
Computers (SBC) and CoProcessors (COP) form the space segment.
The RBC remains on the REXUS rocket for the whole duration of the flight.
Its main tasks are power control for the Spaceseeds (see
Sec.~\ref{sec:power-switch}) as well as telecommand and telemetry forwarding
between the REXUS rocket uplink provided by the REXUS service module and the
Spaceseeds (further discussed in Sec.~\ref{sec:up-downlink}).

The two Spaceseeds are the free-falling units (FFU) that will be ejected at
trajectory apex of roughly \SI{80}{\km},
autonomously control their descent and landing and will then later be
recovered to extract the detailed protocol stored on the
redundant on-board flash storage.
(The ejection is timed by the REXUS service module and is therefore not
a task of either of our systems.)
A Spaceseed has two CPUs, the SBC and the COP.
The SBC is the main computing unit responsible for data acquisition,
communication, the flight controller and data logging;
the COP is responsible for interfacing with additional periphery, such as power
and thermal management.
This split allowed software development to be more focused on their respective
sub-task and made it possible to develop two separate electronic boards which
can undergo development revisions independently, thus a multi-board design for
the board computer allowed for much faster design iterations while also helping
to work around space constraints.

The ground segment consists of our ground station which interfaces with the
uplink to the REXUS service module,
the Iridium network for receiving direct downlinks from the Seeds and
a recovery device which helps locate the landed seeds, which interfaces via
LoRa.
It collects and displays the received data and allows commanding the system
before ejection.
Displaying live data is important as the FFU's GPS position is required for
successful recovery. 

	\section{Electronics}
	\begin{figure}[h]
	\centering
	\includegraphics[width=\columnwidth]{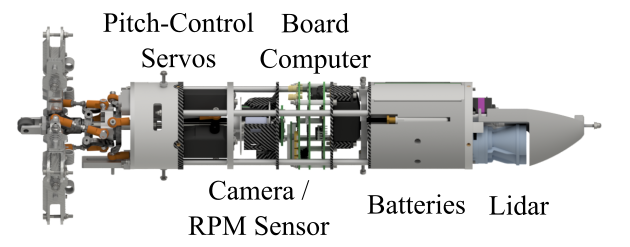}
	\caption{Spaceseed Component Structure}
	\label{fig:seed-no-hull}
\end{figure}

\subsection{Mechanical Challanges}

As a rough outline for maximum accelerations we build on the experience from
Daedalus 1, where one of the three probes experienced forces in excess of
\SI{115}{g} over \SI{20}{\second}.
This occurred because the Spaceseed entered flat spin, a flight configuration
where the main body is not oriented vertical as planned but horizontal, while
rapidly spinning around the z-axis.

And although the new flight controller should prevent another flat spin, the
system should still be robust against such situations.
Together with the accelerations and vibrations on board the rocket and later
during atmospheric entry, this results in extreme mechanical stresses for all
components and especially connectors are at a high risk of loose contacts or
complete failure.

\begin{figure}[h]
	\centering
	\includegraphics[width=0.6\columnwidth]{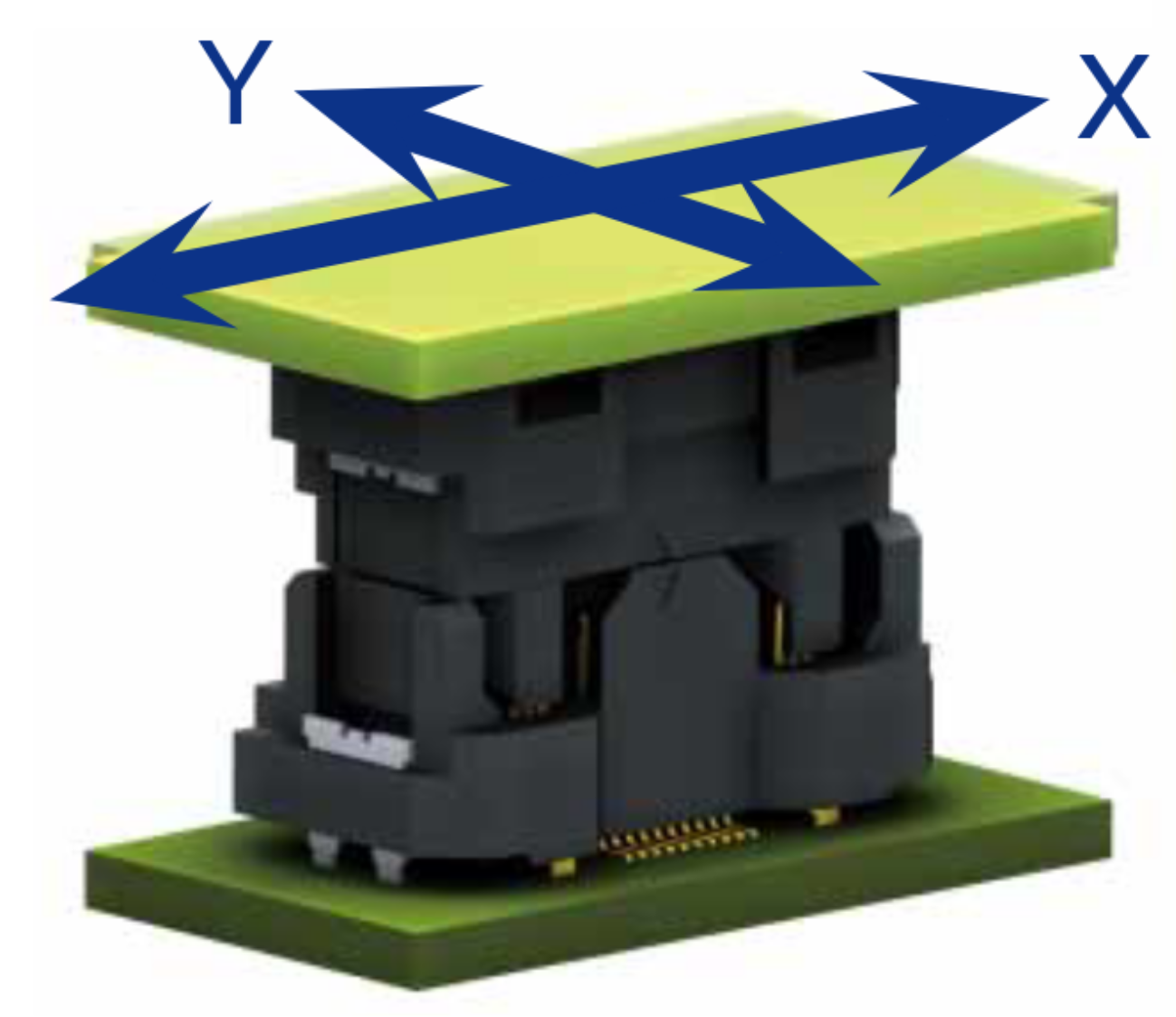}
	\caption{\copyright Hirose Electric Co., Ltd.\\FX23L Connection \cite{fx23l}}
	\label{fig:fx23l}
\end{figure}

For mechanical robustness it was chosen to use a special construction technique
with mechanically decoupled connectors to build a robust multi-board flight
computer.
Specifically Hirose FX23L connectors have the ability to float, i.e. to still
make shock- and vibration-proof connections, even if misaligned by as much as
$\pm$ \SI{0.6}{\milli\meter} (see Fig.~\ref{fig:fx23l}).

\begin{figure}[h]
	\centering
	\includegraphics[width=\columnwidth]{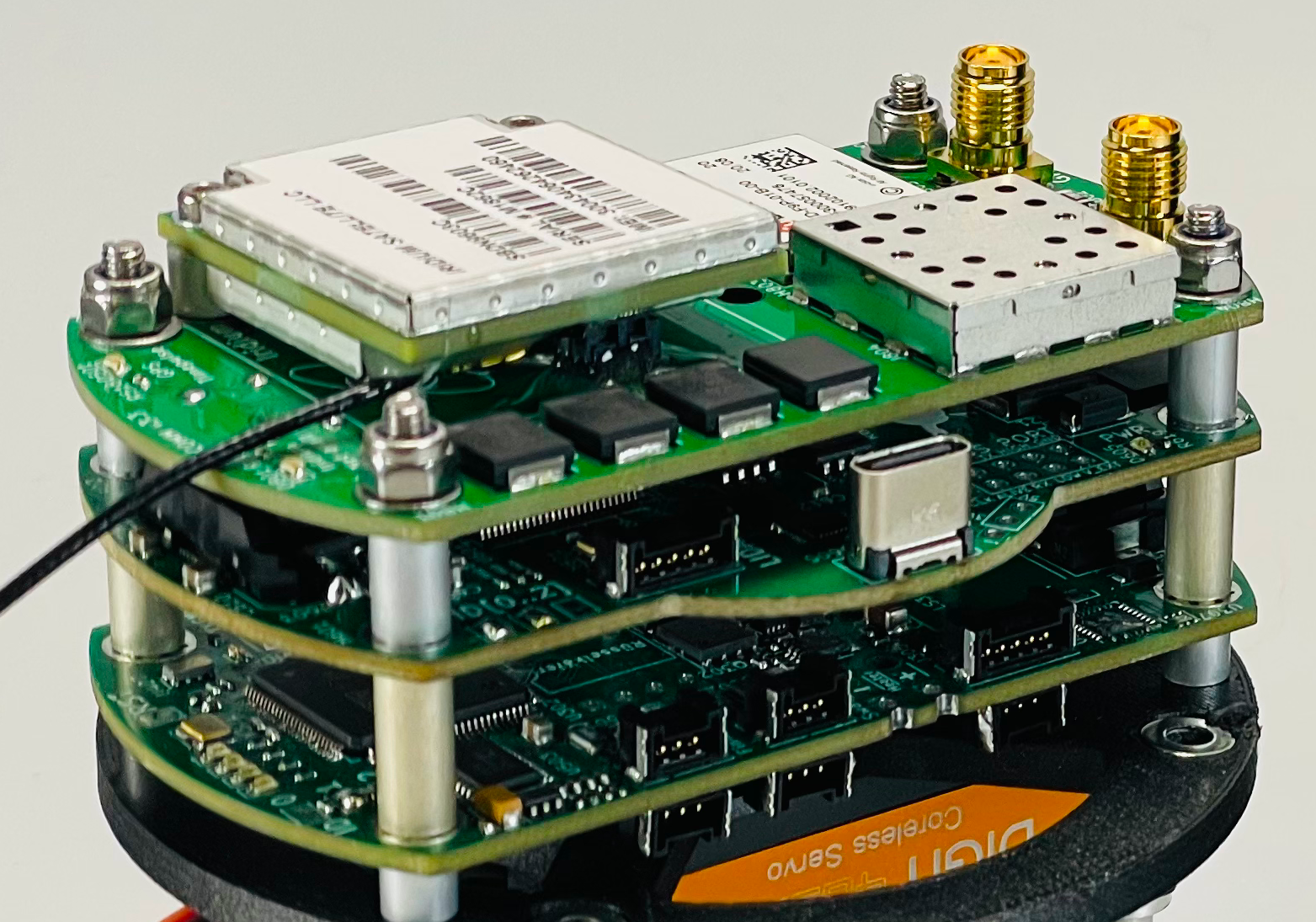}
	\caption{Complete Seed PCB Stack}
	\label{fig:pcb-stack}
\end{figure}

Additionally, each solder connection of wires to PCBs uses two solder holes and
a mechanical hole to provide stress relief.

\subsection{Redundant Sensor Configuration}

The maximum acceleration during the \D1 flat spin could not be determined as the
measurement surpassed the maximum value of \SI{115}{g} that the accelerometer
could measure.
That is why \D2 carries two accelerometers per Spaceseed, where the first is
quite precise and used as input for the flight controller while a second
high-load accelerometer provides measurements up to \SI{400}{g} and could also
be useful as a second reference during result analysis.

A second critical sensor is the barometer because it is the main reference for
determining both height and vertical speed, as GPS won't be available in high
altitudes.
Again two sensors are used, where the first obtains a precise accuracy of
\SI{2.5}{\milli\bar} over the range from \SIrange{10}{1200}{\milli\bar} while
the second has lower accuracy but can reach down to \SI{0}{\milli\bar},
which is useful because much of the flight is performed in very low atmosphere.

\subsection{Battery Choice}
The battery chemistry chosen for the experiment is \ch{Li/SO2}, specifically
SAFT primary cells of type LO 35 SX, because of their non-flammability, big
temperature range, high peak current and the excellent energy density.

\begin{table}[h]
	\small
    \centering
    \caption{\ch{Li/SO2}-cell characteristics \cite{lo35sx}}
    \label{tab:liso2}
    \begin{tabular}{l r}
        \toprule
        Self-Discharge & $\SI{>3}{\percent \per year}$\\
        Storage Temperature & \SIrange{-60}{85}{\celsius}\\
        Operating Temperature & \SIrange{-60}{70}{\celsius}\\
        Capacity & \SI{2.2}{\ampere\hour}\\
        Voltage (at \SI{250}{\milli\ampere} and \SI{20}{\celsius}) & \SI{2.8}{\volt}\\
        Current (cont) & \SI{2}{\ampere}\\
        Current (peak) & \SI{5}{\ampere}\\
        Weight & \SI{30}{\gram}\\
        $\implies$ Energy Density & \SI{0.74}{\mega\joule\per\kilo\gram}\\
        \bottomrule
    \end{tabular}
\end{table}

The cells are employed in a 3S2P configuration, leading to a nominal voltage of
\SI{8.4}{\volt} at a capacity of \SI{4.4}{\ampere\hour}.
For redundancy purposes each individual battery strings should suffice for a
successful mission. In the case of failure of a single string it will be
automatically disconnected using the circuit described in Sec.~\ref{sec:power-switch}.

\subsection{Power Topology}
\label{sec:power-switch}
The Spaceseeds can be powered from \SIrange{6}{40}{\volt} via three inputs.
Two inputs are used for the batteries, while the third is supplied by the Rocket
Board Computer which relays the \SI{28}{\volt} supply from the REXUS Service
Module (RXSM).

Upon Spaceseed ejection the RXSM connection is severed and the experiment shall
seamlessly switch from the external power supply to the internal batteries.
In its simplest form this can be achieved with diode OR-ing, where each supply
has a Schottky diode, such that the highest supply voltage will always power
experiment.

\begin{figure}[h]
	\centering
	\begin{circuitikz}
		\draw (0,-1) to[empty Schottky diode] (2,-1);

		\node at (4,0) [pigfete, rotate=90, yscale=-1](q1){};
		\node at (5,0) [pigfete, rotate=90](q2){};
		\draw (q1.D) to (3,0);
		\draw (q2.D) to (6,0);
		\draw (q1.G) to (q2.G);

		\node at (1,-2.5){$P=V_{\text{DROP}}~ I_{\text{LOAD}}$};
		\node at (4.5,-2.5){$P=2~ {I_{\text{LOAD}}}^2~ R_{\text{DS(ON)}}$};

		\draw (4.5,-0.77) to[short,*-] (4.5,-1.25);
		\draw (3.5,0) to[short,*-] (3.5,-1.25);
		\draw (5.5,0) to[short,*-] (5.5,-1.25);
		\draw (3.25,-1.25) -- (5.75,-1.25) -- (5.75,-2) -- (3.25,-2) -- cycle;

		\node at (4.5,-1.625)[align=center]{Ideal Diode\\Controller};
	\end{circuitikz}
	\caption{Comparison of Power Loss for Traditional Schottky Diode vs Ideal MOSFET Diode}
	\label{fig:diode-comparison}
\end{figure}

This has the downside of a voltage drop of roughly \SI{400}{\milli\volt}
across the diode which causes additional power dissipation.
Instead, an ideal diode circuit was implemented that uses two back-to-back
MOSFETs.
This way the power dissipation is only dependent on the drain-source on
resistance $R_{DS(ON)}$, which can be as low as $\SI{5}{\milli\ohm}$.
Two FETs are needed to both switch the supply current and block any return
charging currents.

\begin{figure*}[h]
	\centering
	\includegraphics[width=\textwidth]{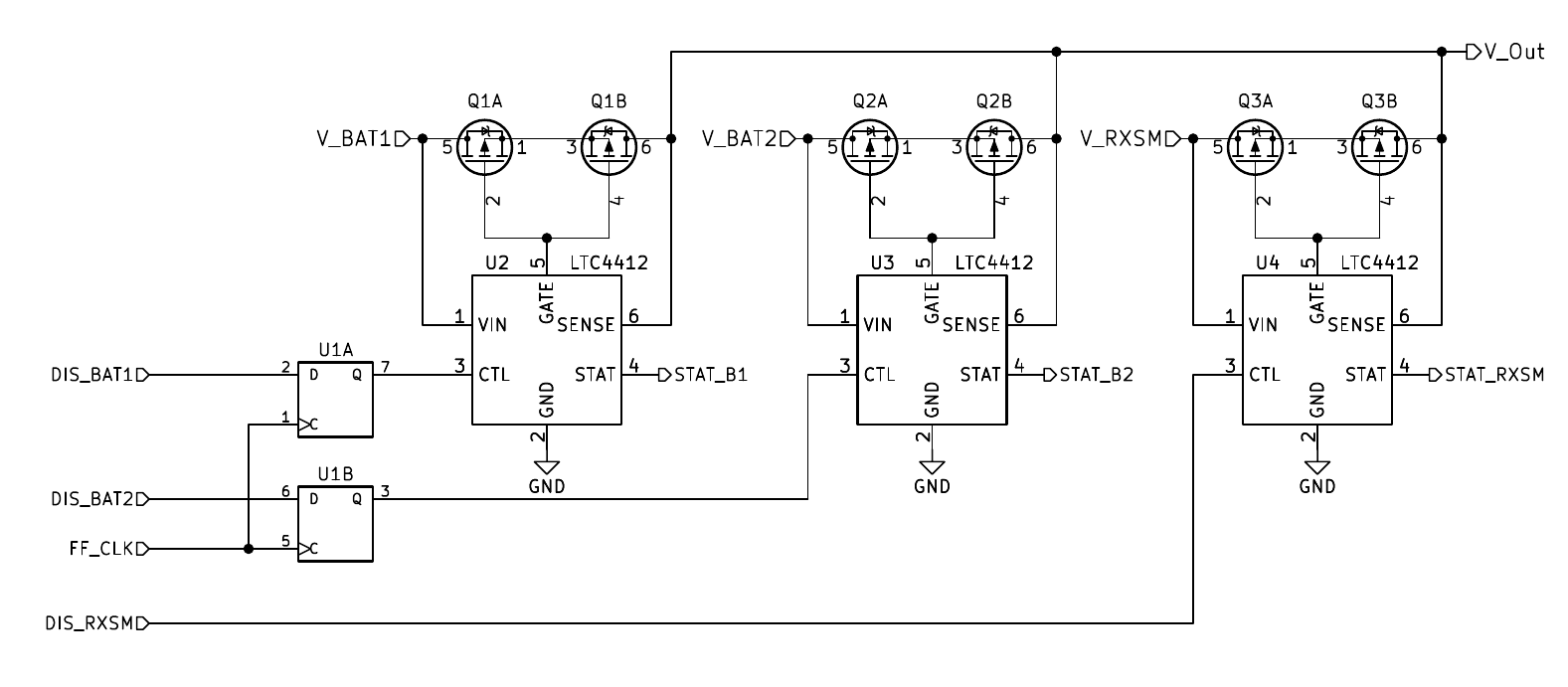}
	\caption{Schematic of the Power Switch Circuit}
	\label{fig:power-circuit}
\end{figure*}

To switch the MOSFETs, an ideal diode controller is employed which compares the
voltages across the ideal diode and switches the gates if needed.
By using a small hysteresis of \SI{20}{\milli\volt} for the comparator both
batteries can supply the experiment at the same time if their voltages are close
enough.

An additional advantage is that the controller can be used by the COP to
completely switch off a power supply if desired.
This is required by REXUS guidelines for the phase of ``radio silence'', in
which any currents inside the experiment are strictly forbidden.
For example during the critical stage of the rocket motor integration ``radio
silence'' is employed, as ground currents across the rocket hull may trigger
accidental ignition.

For this we use two D flip-flops as seen in Fig.~\ref{fig:power-circuit},
which can be set by the SBC using \texttt{DIS\_BAT1}, \texttt{DIS\_BAT2} and a
rising edge on \texttt{FF\_CLK}.
The flip-flops are always powered by their respective battery, so to enter
radio silence the following procedure takes place:

\begin{itemize}
	\item Experiment is supplied by \texttt{V\_BAT1}, \texttt{V\_BAT2} and the
		rocket supply \texttt{V\_RXSM}.
	\item Radio silence is requested, the COP sets both flip-flops.
	\item The ideal diode controllers for the batteries switch off, the
		experiment is only supplied via \texttt{V\_RXSM}.
	\item The RBC cuts the \texttt{V\_RXSM} supply, the Spaceseeds completely
		switch off while the flip-flops prevent the circuit from waking up via
		battery power.
	\item Power to the RBC is cut and the experiment is completely switched off.
	\item When reapplying external power the RBC boots, power is applied to the
		Spaceseed and the COP reactivates both batteries.
		Now the system is again armed for ejection.
\end{itemize}
	
\subsection{Recovery Strategy}

The recovery of the fallen Spaceseeds is a secondary objective, since for the performance evaluation
of the autorotation system as parachute alternative the wireless data may be sufficient.
Still, successful recovery has a high priority so redundant systems were devised.

\subsubsection{GPS Positions}

Each Spaceseed has a u-blox ZED-F9P GNSS module to determine descent speed and
position with an accuracy of \SI{0.01}{\meter} + \SI{1}{ppm} CEP at an update
rate of \SI{20}{\hertz}.
The position estimate is used as flight controller input, but also transmitted
over the Iridium network to the ground station backend server which can display
a landing prediction on a ground station frontend.

Although successfully used on \D1 this setup relies on the availability of two
satellite networks, GPS and Iridium, as well as a working power system.

\subsubsection{LoRa}

If only a coarse location of the unit is known,
possibly through GPS positions transmitted some time before the landing via
Iridium or trajectory extrapolation,
the recovery crew needs assistance in location the FFU,
which may be hidden in the snow or a forest.

This recovery mechanism works by having the landed units repeatedly broadcast
their GPS position if available - or dummy data otherwise - over a long range
(LoRa) radio module, which provides multiple kilometers of range albeit at low
data-rates.

A purpose-built recovery device will receive these messages transmitted by the
Spaceseeds if they are in range and display the received data but also the
received signal power.
By connecting a directional antenna to the device and turning on the spot,
the direction of the transmitter on-board the Spaceseed can be determined.
Repeatedly perform this maneuver and moving in the indicated direction
should lead the crew to the FFU.

\subsubsection{RECCO Reflectors}

Additionally, a passive recovery system is deployed which was originally devised
for avalanche search and rescue (SAR) operations.
It uses a passive reflector that is usually built into the garments worn by
users and which is lightweight (\SI{4}{\gram}) and small, such that multiple
reflectors can be attached to a single Spaceseed.
The active detector is either a hand-held device or attached to a SAR helicopter
and has a range of \SI{200}{\meter} through air and \SI{30}{\meter} through
snow. 

This system has the advantage that it does not rely on being powered or having a
working communication downlink, but due to its limited range should only be
considered as fallback.

	\section{Communication Channels}
	The \D2 experiment consists of multiple components that need to communicate with
very different network requirements in concern to reliability, latency and
bandwidth.
\begin{itemize}
  \item During testing we need to transmit all data at full bandwidth if a
  wired connection is available, but also support streaming a subset of the 
  data wirelessly if the testes required it. An uplink must also be supported.
  \item The COP and SBC need high bandwidth and reliable communication to
  exchange telemetry and control actuators.
  \item The communication between the RBC and SBC must be reliable
  and correctly handle the disconnect during ejection.
  \item Communication between the RBC and the ground station is low-bandwidth
  but must be reliable for commanding and telemetry reporting before lift-off.
  \item Communication between the Seeds and the ground station must be reliable
  for reporting the GPS positions and ideally as much data as possible in case
  recovery is not possible.
  \item We need a way to help the recovery crew find the FFUs.
\end{itemize}


We addressed these challenges with the following design, which will be discussed
in more detail in the rest of this section.
\begin{itemize}
  \item Use a publish/subscribe messaging system between the space components.
  \item Use MAVLink for communication via the REXUS service module and for
  recovery.
  \item Use Iridium for the direct downlink from the Spaceseeds. 
\end{itemize}

\subsection{Publish/Subscribe Messaging}
Our software uses the real-time operating system and middleware
RODOS\footnote{Current development happens at \url{https://gitlab.com/rodos/rodos}}
\cite{MontenegroDannemann2009RODOS}
which was originally designed for use in satellites.
It includes a publish/subscribe message passing system.
This allows us to exchange product types (C structs) between different software
components, independent of whether they are located on the same or a different
computing node.
Decoupling components this way increases flexibility and reusability but
necessarily increases complexity.

This is used extensively in our space segment to communicate between software
components on the same and different nodes.\todo{die info find ich muss noch mit
rein, aber irgendwie muss man das besser unterbringen}
To facilitate communication between different nodes, multiple transport
protocols (\texttt{LinkInterface} in RODOS terms) may be used.
Between the SBC and COP we use a simple UART connection because of its
simplicity and reliability.
The RBC and the two Spaceseeds communicate with a CAN bus.
It was chosen because of its ability to connect more than two nodes and the
availability of an ACK system.
We learned however that our design here resulted in increased complexity during
ejection as bus arbitration is no longer possible for participants without a
termination afterwards.

\subsection{Up- and Downlink}
\label{sec:up-downlink}
As mentioned above,
two protocols are used for the communication between the space and the ground
segment to balance the different constraints and needs on each channel.
They shall be discussed here in more detail.

\subsubsection{MAVLink}
MAVLink\footnote{MAVLink's project website is at the time of writing available
at \url{https://mavlink.io/en/}} (Micro Air Vehicle Link)
is the protocol we choose for the down- and uplink during testing
and via the REXUS service module.
It was originally designed for use with drones over wireless channels and
therefore already provides message framing, routing information
(sender and destination), error detection, message authentication
and replay protection.

It is implemented as a code generator which reads a message definition
and is able to emit package generation and parsing code in multiple languages,
including C++ (used for embedded development) and Java (used for the ground
station, see Sec.~\ref{sec:ground-station} for more info).
This reduces development time and increases reliability,
as no parsing code has to be written manually,
which eliminates a large field of possible errors.

The error detection and message authentication are not strictly required for
the wired testing connection.
To reduce the number of different software systems
and thereby keep complexity limited,
we decided to use MAVLink here anyway.

\todo[inline]{MAVLink replay protection?}

\todo[inline]{MAVLink routing?}


\subsubsection{LoRa}
Both Spaceseeds carry a LoRa transceiver,
which is used to provide a wireless testing connection
and an additional recovery mechanism.
LoRa is a proprietary IoT modulation scheme and protocol optimized for a
long range by use of Chirped Spread Spectrum (CSS) modulation,
but only offering a low data rate as a trade-off.
It is currently developed by Semtech.

\subsubsection{Seed Downlink: Iridium}
Because of our experiences with \D1 we choose to again use the Iridium network
as our Spaceseed downlink.
Due to the high cost of satellite data transmissions however we do not use
MAVLink as it would introduce overhead like including routing information
twice per packet.
Instead, we transmit raw C structs which increases complexity as for example
endianness and message versioning needs to be considered manually instead
of letting the software automatically handle it.
This choice allows us to maximize transmission of science and position data.
The latter is crucial, as it is required for recovery of the Spaceseeds
with their flash memories, which hold full-rate science data.

The Iridium satellite constellation consists of 66 cross-linked low earth orbit
satellites at an altitude of $\SI{780}{\kilo\meter}$.
This provides global network (including polar coverage) that guarantees a low
possibility of signal loss and highly reliable communication.
By using the Iridium Short Burst Data (SBD) service each Spaceseed will stream
their position and some key telemetry over a passive quadrifilar helix antenna
to the network.
This data is then transmitted over TCP to a backend server that is part of the
\D2 ground station infrastructure to enable successful recovery.

\subsection{Ground Station}
\label{sec:ground-station}
The data arrives at ground station via different channels in different formats
(MAVLink and C structs from Iridium).
Its main job is receiving, storing and presenting this data to the user.
To achieve this, the data is unpacked and transformed into a common,
self-describing format (JSON) for storage and further processing.
It must also be able to command the system during testing
and while it is on the launch pad.

This is achieved by the highly modular and reusable ground station framework
\emph{Telestion}\footnote{The software is open source and available online at
\url{https://github.com/wuespace} in the \texttt{telestion-*} repositories,
the project specific code is available at
\url{https://github.com/wuespace/telestion-project-daedalus2}}.
It separates the backend,
where incoming and outgoing data is processed and
stored for future analysis and review,
from the frontend which displays received telemetry
and also provides the telecommand interface.

This separation allows multiple operators to command and inspect the system
from multiple computers.
Additionally, it also increases reliability
as the amount of code in the critical processes (the backend) is reduced,
and the components responsible for communication between front- and backend
as well as storage are well tested,
as they can be and have successfully been used for other projects as well.

\todo[inline]{Screenshot of GS?}

	\section{Pre-Flight Evaluation}
	To verify the system design before the actual flight on the REXUS rocket
multiple evaluation tests were performed:

\begin{figure}
    \centering
    \includegraphics[width=0.9\columnwidth]{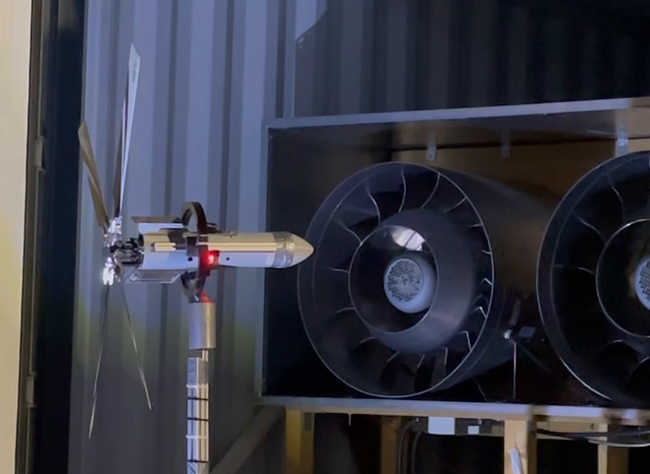}
    \caption{Gimbal Mounted Spaceseed in Wind Tunnel}
    \label{fig:wind-tunnel}
\end{figure}

\begin{itemize}
    \item Bench tests to verify the whole communication chain end-to-end
    \item Location tests using the LoRa Recovery Device
    \item Wind tunnel tests to simulate flight conditions
\end{itemize}

The wind tunnel tests in particular provided valuable insights as it forced
all subteams (in particular the mechanical, embedded software, electronics and
simulation team) to work together and synchronize their progress.

\subsection{Power System Verification}

In a standard test environment the tunnel produces constant wind speeds of
roughly \SI{15}{\meter\per\second} while the flight controller adjusts the blade
pitch of the rotor to reach a target rotor rotation speed.
After a pre-programmed time interval a new target speed is chosen to observe
the step response of the control algorithm.
All sensor readings as well as various internal states would be saved to the
flash storage for later detailed analysis with a rate of \SI{250}{\hertz}, just
like in a real flight.

\begin{figure}[t]
    \includegraphics{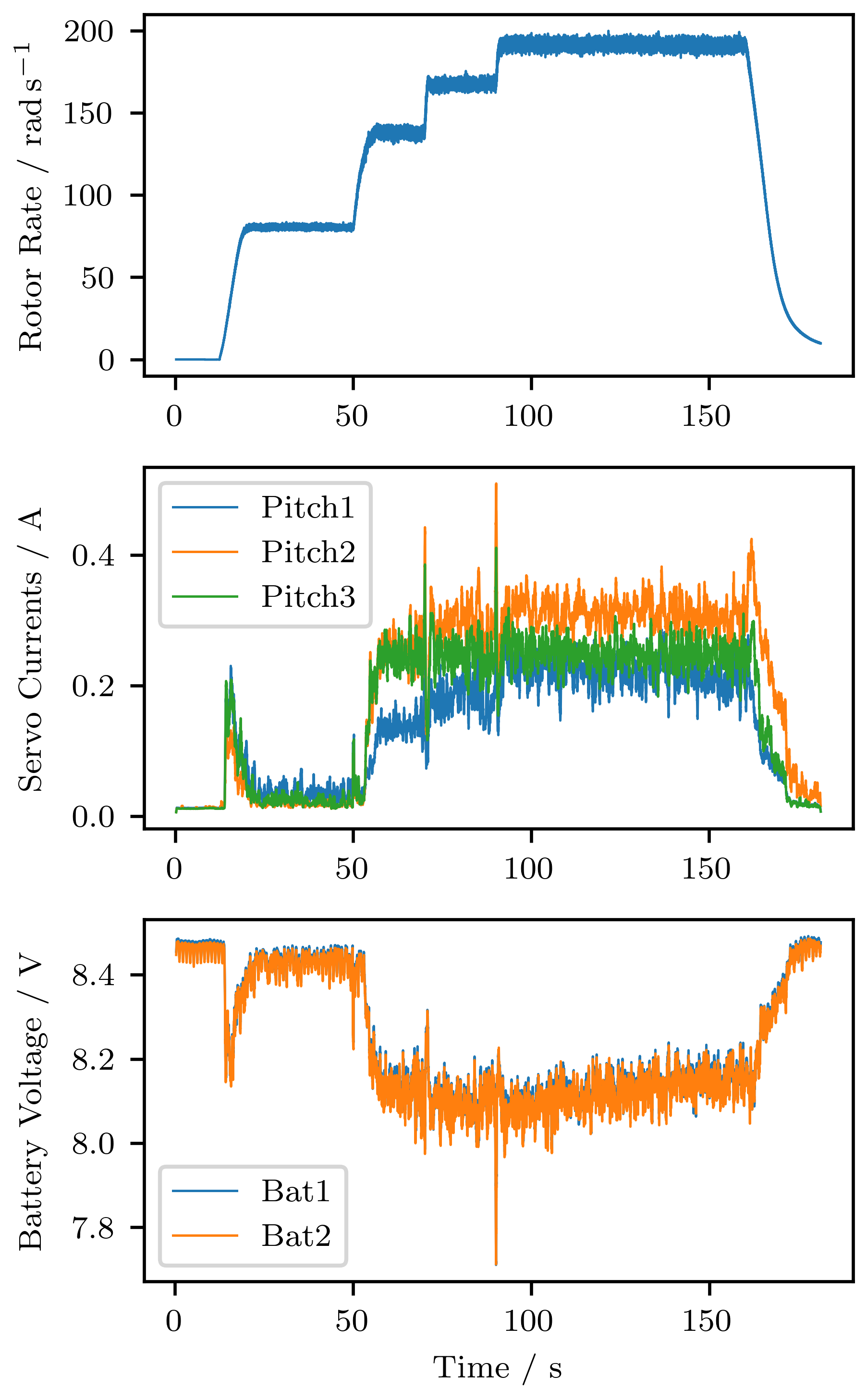}
    \caption{Electrical Performance}
    \label{fig:elec-perf}
\end{figure}

One such analysis can be seen in Fig.~\ref{fig:elec-perf}, where the power
system was evaluated with special focus on the current draw of the three servos
that control the blade pitch.
This showed that even under maximum rotor speeds the current was well within
design limits, both battery strings were utilized equally, and no critical
voltage drops occurred during pitch changes.

\subsection{Tachometer Verification}

Another critical test involved the verification of the tachometer that measures
the rotor rotation rate.
This sensor was built from scratch using dedicated logic gates and low-level
microcontroller periphery, as well as compensation factors in software.
Naturally, the measurements needed to be independently verified before the sensor
output could be trusted.

\begin{figure}[t]
    \includegraphics{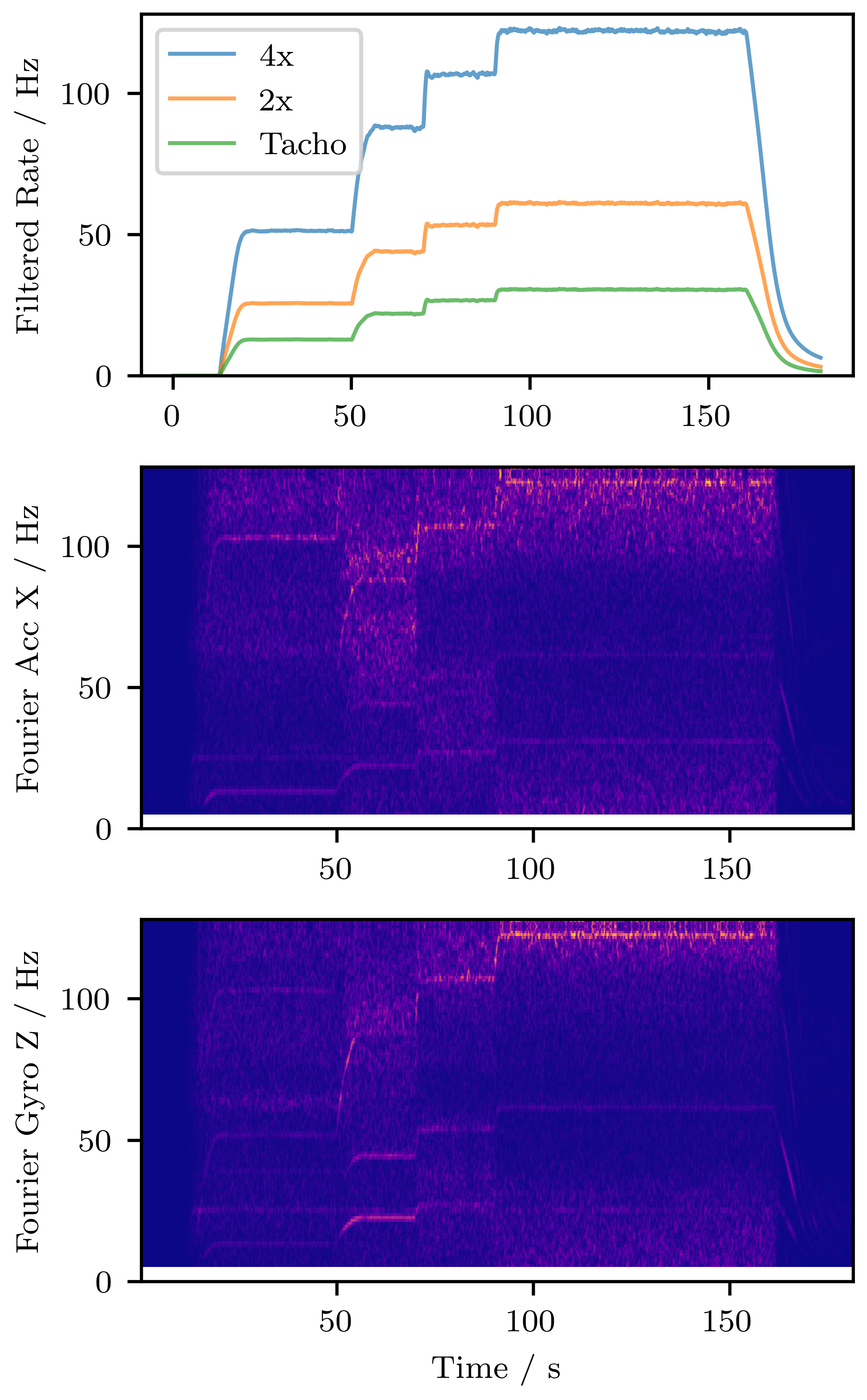}
    \caption{Comparison of Tachometer Output and Frequency of Accelerometer/Gyroscope Vibrations}
    \label{fig:tacho-valid}
\end{figure}

For this the accelerometer and gyroscope readings were used, as the rotor will
also induce vibrations corresponding to the rotation rate.
These frequencies can be determined by doing a Fourier analysis on the measurements
as seen in Fig.~\ref{fig:tacho-valid}.
Indeed, three distinct waveforms can be seen that correspond exactly to the
readings from the tachometer as well as the second and fourth harmonics.
The fourth harmonic is the most dominant, which makes sense as the rotor
consists of four blades, so most of the vibrations will have a frequency four
times higher than the full rotor rotation.

	\pagebreak

	\section{Conclusion}
	This proposed structure, from electronics over communication to system topology, proved valuable
and delivered good performance in pre-flight tests such as mechanical shaker tests and flight
simulations in a wind-tunnel.
Of course, it is impossible to predict all aspects of a real flight, but the system appears
robust and will be used as flight hardware in March 2023.

	\section{Acknowledgements}
	There are many people and institutions who have supported us in various ways
during the project and who deserve to be mentioned here.
As far as institutions are concerned, we would like to start with thanking the
German Aerospace Center DLR, the Swedish National Aerospace National Space
Agency SNSA and the European Space Agency ESA as the main organizers of
the REXUS/BEXUS program.
In addition, the campaign would not be possible without the Swedish Space
Corporation SSC.
The Center for Applied Space Technology and Microgravity ZARM also deserves a
special Thanks for their continuous support to the team in making improvements
and successfully completing the project cycle.
Furthermore, we would like to thank the Julius-Maximilians-University Würzburg,
in particular the Chair for Aerospace Information Technology [10], represented
by Prof. Hakan Kayal.
We also express our gratitude to Prof. Sergio Montenegro, who let us use crucial
infrastructure like the wind tunnel, an electronics laboratory or his SKITH
boards for development during the chip crisis.

The Daedalus 2 team is grateful to all sponsors who made the project possible:
Atomstreet, Mouser Electronics, Carbonteam Shop, Hacker Brushless Motors,
MathWorks, Siemens, ublox, VDI, Breunig Aerospace, Airbus, Ansys, Hirsch KG, PCB
Arts, Vogel Stiftung, MÄDLER, WÜRTH ELEKTRONIK, PartsBox, molex, LRT Automotive,
ZfT, Iridium, TELESPAZIO and ARCTIC.
An overview of all helpers and sponsors can be found in
\cite{wuespace-supporters}.

	\printbibliography

@InProceedings{D1ESAPAC,
  author    = {Riegler, Clemens and Angelov, Ivaylo and Kohmann, Florian and Neumann, Tobias and Bilican, Abdurrahman and Hofmann, Kai and Pielucha, Jessica and Böhm, Alexander and Fischbach, Barbara and Appelt, Tim and Willand, Lisa and Wizemann, Oliver and Menninger, Sarah and von Pichowski, Jan and Staus, Jonas and Hemmelmann, Erik and Seisl, Sebastian and Fröhlich, Christoph and Plausonig, Christian and Rath, Reinhard},
  booktitle = {23rd ESA PAC},
  title     = {{PROJECT DAEDALUS, ROTOR CONTROLLED DESCENT AND LANDING ON REXUS23}},
  year      = {2019},
  month     = jun,
  url       = {https://www.researchgate.net/publication/337819783_PROJECT_DAEDALUS_ROTOR_CONTROLLED_DESCENT_AND_LANDING_ON_REXUS23},
}

@Manual{lo35sx,
  title        = {{3.0V} Primary lithium-sulfur dioxide {(\ch{Li-SO_2})}},
  month        = {09},
  organization = {SAFT},
  year         = {2008},
  number       = {LO35SX},
}

@Manual{fx23l,
  title        = {High-speed transmission {\SI{0.5}{\milli\meter}} pitch board-to-board connection floating connectors},
  month        = {07},
  organization = {Hirose Electric Co., Ltd.},
  year         = {2022},
  number       = {FX23/FX23L Series},
  url          = {https://www.hirose.com/product/document?clcode=&productname=&series=FX23L&documenttype=Catalog&lang=en&documentid=D141086_en},
}

@Article{MontenegroDannemann2009RODOS,
  author  = {Montenegro, Sergio and Dannemann, Frank},
  journal = {DASIA 2009 - DAta Systems in Aerospace},
  title   = {{RODOS} - real time kernel design for dependability},
  year    = {2009},
  pages   = {66},
  volume  = {669},
  comment = {Quelle: von Monty per E-Mail erhalten},
}

@inproceedings{D2SSEA,
  author  = {Mehringer, Johanna and Werner, Lennart and Riegler, Clemens and Dunschen, Frederik},
  year    = {2022},
  month   = {04},
  pages   = {},
  title   = {Suborbital autorotation landing demonstrator on REXUS 29},
  doi     = {10.5821/conference-9788419184405.039},
  url     = {https://upcommons.upc.edu/bitstream/handle/2117/369264/SSEA_2022_102.pdf}
}

@online{wuespace-supporters,
  title   = {Supporters, Online Presence of WüSpace e.~V.},
  url     = {https://www.wuespace.de/supporters/}
}
\end{document}